\documentclass[a4paper]{jpconf}
\usepackage[dvips]{graphicx}
\usepackage{times}
\usepackage{amsmath}
\usepackage{braket}
\usepackage{xcolor}
\usepackage{orcidlink}
\usepackage{hyperref}
\usepackage{color,soul}
\hypersetup{
  colorlinks=true,
  urlcolor=blue,
  linkcolor=red,
  citecolor=blue
}
\begin{document}
\title{Propagation of UHE cosmic rays in the light of f(R) gravity power-law model}

\author{Swaraj Pratim Sarmah}
\ead{swarajpratimsarmah16@gmail.com}

\address{Department of Physics, Dibrugarh University, Dibrugarh 786004, 
Assam, India}

\begin{abstract}
While the origins of ultra-high energy (UHE) cosmic rays (CRs) remain shrouded in uncertainty, several important milestones have been reached in recent years in the experimental study of CRs with energy above $10^{18}$ eV. Within the vast expanse of intergalactic space, turbulent magnetic fields (TMFs) are believed to pervade, and these fields could exert a significant influence on the journey of UHECRs across the expanding Universe, which is currently undergoing acceleration. Thus, it is imperative to incorporate these considerations into our theoretical framework to gain a deeper understanding of the empirical observations related to UHECRs. In light of this, our research delves into the impact of UHE particle diffusion in the presence of TMFs, all within the context of the f(R) gravity power-law model. Based on this f(R) model, we explore the diffusive behavior of UHECR protons, particularly focusing on their density enhancement throughout their propagation and their energy spectrum. We found that the f(R) gravity model considered here plays an effective role in the propagation of CRs and the results have lain within our range of interest. Also, we compare our results for flux with observational data like the Pierre Auger Observatory (PAO) and Telescope Array (TA). 
\end{abstract}


\section{Introduction}

Cosmic rays (CRs) are ionizing particles that come from the outside of our solar system. V.\ F.\ Hess discovered them in 1912 \cite{Hess}, which was a breakthrough for modern physics. However, even after more than a hundred and ten years, many questions about CRs remain unanswered \cite{harari,molerach,berezinskyGK}. For example, we do not know how they are created, accelerated, and propagated, especially when they have very high energies ($E\geq 0.1$ EeV, where $1$ EeV = $10^{18}$ eV) \cite{berezinskyGG, Bhattacharjee, Olinto}. We think that the sources of lower energy CRs ($E \leq 0.1$ EeV) are galactic and are related to supernova explosions \cite{s.mollerach}, but the sources of higher energy CRs ($\sim 1$ EeV and above) are probably outside our galaxy i.e. extragalactic, and could be linked to gamma-ray ($\gamma$-ray) bursts.

The motion of a charged particle in an arbitrary magnetic field is contingent upon the distance it covers relative to the scattering length, $\l_{D} =3D/c$, where $D$ is the diffusion coefficient \cite{Supanitsky}. If the particle's journey is significantly shorter than the scattering length, its propagation exhibits ballistic characteristics. Conversely, if the distance covered greatly exceeds the scattering length, the propagation is diffusive. Incorporating an extragalactic Turbulent Magnetic Field (TMF) and accounting for a finite source density in UHECR propagation studies may lead to a low-energy magnetic horizon effect. This could reconcile observations with a higher spectral index \cite{s.mollerach, mollerach2013}, aligning more closely with predictions from diffusive shock acceleration.

In view of the limitations of General Relativity (GR) such as 
expanding acceleration and the missing mass of the Universe, we have preferred to choose modified theories of gravity (MTGs) to study the characteristics of the propagation of UHECRs. 
We have discussed the TMF and the required mathematical expressions including the cosmological model in section \ref{II}. 
The results of our work are discussed in section \ref{III}. 
At last, we have summarised and concluded our work in section \ref{IV}.

\section{Propagation of UHECR in turbulent magnetic field} \label{II}
One of the difficulties in modeling the extragalactic magnetic field is the lack of sufficient observations to constrain them due to their exact strength being uncertain, and they may vary depending on the spatial region under consideration.
We will focus on the propagation of the CRs, which is a complex problem. The rotation measure of the cosmic microwave background (CMB) polarised sources indicates that a strong magnetic field exists in the local supercluster, with an estimated strength of 0.3 to 2 $\mu$G. This magnetic field is the most relevant
one since it affects the CRs from the nearest sources which can be reached at the earth's surface. We will ignore the larger-scale variations
from  voids and filaments. We will assume that the CRs propagate
in an isotropic, homogenous, turbulent extragalactic magnetic
field, which is a simplification. Such magnetic fields can be
characterised by the RMS amplitude of magnetic field B and the coherence length,
$l_c$, which is the maximum distance between two points where
the magnetic fields are correlated. The rms strength of the magnetic fields can range from 1-100 nG, and the
coherence length can range from 0.01-1 Mpc.

We can define the Larmor radius with energy $E$ and charge $Ze$ in a TMF with strength $B$ as
\begin{equation}\label{larmor}
r_L = \frac{E}{ZeB} \simeq 1.1 \frac{E~ . \textrm{nG}}{ZB ~. \textrm{EeV}}\;\textrm{Mpc}.
\end{equation}
Another important quantity for studying the  charged particles' diffusion is critical energy, given as
\begin{equation}\label{cri_energy}
E_c = ZeBl_c \simeq 0.9 Z\, \frac{B}{\textrm{nG}}\, \frac{l_c}{\textrm{Mpc}}\;\textrm{EeV}.
\end{equation}
This energy separates the resonant diffusion regime at low energies ($<E_c$) from the non-resonant regime at high energies ($>E_c$). Harari et al. proposed a simulations for propagation of CRs that is fitted with the diffusion coefficient $D(E)$ \cite{harari}, which is
\begin{equation}\label{diff_coeff}
D(E) \simeq \frac{c\,l_c}{3}\left[4 \left(\frac{E}{E_c} \right)^2 + a_I \left(\frac{E}{E_c} \right) + a_L \left(\frac{E}{E_c} \right)^{2-m}   \right],
\end{equation}
where $m=5/3$ denotes spectral index, $a_I \approx 0.9$ and $a_L \approx 0.23$ are two coefficients with the Kolmogorov spectrum.

The UHE particles that propagate in expanding Universe at a source distance ${\bf x}_s$ obey the diffusion equation in the diffusive regime \cite{berezinkyGre} as
\begin{equation}\label{diff_eqn}
\frac{\partial n}{\partial t} + 3 H(t)\,n - b(E,t)\frac{\partial n}{\partial E}-n\, \frac{\partial n}{\partial E}-\frac{D(E,t)}{a^2(t)}\,\nabla^2 n = \frac{Q_s(E,t)}{a^3(t)}\,\delta^3({\bf x}-{\bf{x}_s}),
\end{equation}
where $H(t)$ denotes the Hubble parameter with the scale factor $a(t)$ dependency,  $n$ represents particle density and the source function $Q_s(E)$  gives at energy $E$ how many
number of particles per unit time emitted.
The general solution of the above equation was derived in
\cite{berezinkyGre} by assuming the particles being protons as
\begin{equation}\label{density}
n(E,r_s)= \int_{0}^{z_{i}} dz\, \bigg | \frac{dt}{dz} \bigg |\, Q(E_g,z)\, \frac{\textrm{exp}\left[-r_s^2/4 \lambda^2\right]}{(4\pi \lambda^2)^{3/2}}\, \frac{dE_g}{dE},
\end{equation}
where $E_g$ is the generation energy at redshift
$z$ of a particle with energy $E$ at redshift $z=0$. $Q(E_g,z)$ is the source function and follows  $ E_g^{-\gamma_g}$ distribution with $\gamma_g$ as the
generation spectral index at the source, and  
\begin{equation}
\lambda^2(E,z)=\int_{0}^{z}dz\, \bigg | \frac{dt}{dz} \bigg |\,(1+z)^2 D(E_\text{\bf g},z).
\end{equation}
$dE_g/dE$ is the rate of energy loss of particles with respect to their energy $E$ at $z=0$ \cite{berezinkyGre, berezinski_four_feat}
The expression for $dE_g/dE$ is taken from Berezinsky et al. \cite{berezinski_four_feat}. The enhancement of density $\xi$, i.e. a measurement of change of density due to diffusion and interaction with cosmic microwave background, can be written as
\begin{equation}\label{enhance}
\xi(E,r_s)=\frac{4\pi r_s^2c\, n(E, r_s)}{\mathcal{L}(E)},
\end{equation} 
here $\mathcal{L}(E)$ depicts spectral emissivity and it follows the power-law relation with the particle energy.
For the energy spectrum of CRs, we calculate the flux of UHECRs and the expression for this quantity can be written as  \cite{ berezinkyGre}
\begin{equation}\label{flux}
J_p (E)=\frac{c}{4\pi}\,\mathcal{L}_0 K \int_{0}^{z_{max}}\!\! dz\, \biggl | \frac{dt}{dz}\biggl |\, (1+z)^m q_{gen}(E_g)\, \frac{dE_g}{dE}.
\end{equation}
 
Now we will consider the f(R) power-law model and its functional form is given by
\begin{equation}
f(R) = \alpha R^\beta
\end{equation}
where $\alpha$ and $\beta$ are model parameters. A detailed discussion of this f(R) gravity model using the palatini formalism and the model parameters
are performed in \cite{gogoi} and the best-fitted value of $\beta$ is found to be 1.4 \cite{gogoi}. The Hubble parameter expression for the f(R) gravity power law model can be written as  \cite{gogoi}
\begin{equation}\label{powerlawhubble}
H(z) = \left[-\,\frac{2 \beta R_0}{3 (3-\beta)^2\, \Omega_{m0}} \Bigl\{(\beta-3)\Omega_{m0}(1+z)^{\frac{3}{\beta}} + 2 (\beta-2)\,\Omega_{r0} (1+z)^{\frac{n+3}{\beta}} \Bigl\}\right]^\frac{1}{2},
\end{equation}

where $\Omega_{m0} \approx 0.31$ \cite{aagnamin} and $\Omega_{r0} \approx 0.000053$ \cite{nakamura} are the matter and radiation density at the present value, and
\begin{equation}
R_0 = -\, \frac{3 (3-\beta)^2 H_0^2\, \Omega_{m0}}{2\beta \left[(\beta-3)\Omega_{m0} + 2 (\beta-2) \Omega_{r0}\right]}.
\end{equation}
is the present Ricci scalar value with $H_0\approx 68.4 ~\text{km}~ \text{s}^{-1}~ \text{Mpc}^{-1}$  and the cosmological time evolution with respect to the redshift can be expressed as,
\begin{equation}\label{dtdz1}
\bigg | \frac{dt}{dz} \bigg | = \frac{1}{1+z}  \left[-\,\frac{2\beta R_0}{3 (3-\beta)^2 \Omega_{m0}} \Bigl\{(\beta-3)\Omega_{m0}(1+z)^{\frac{3}{\beta}} + 2 (\beta-2)\Omega_{r0} (1+z)^{\frac{\beta+3}{\beta}} \Bigl\} \right]^{-\,\frac{1}{2}}\!\!.
\end{equation}
 
\section{Results}  \label{III}
\begin{figure}[t!] 
\centerline{
\includegraphics[scale=0.25]{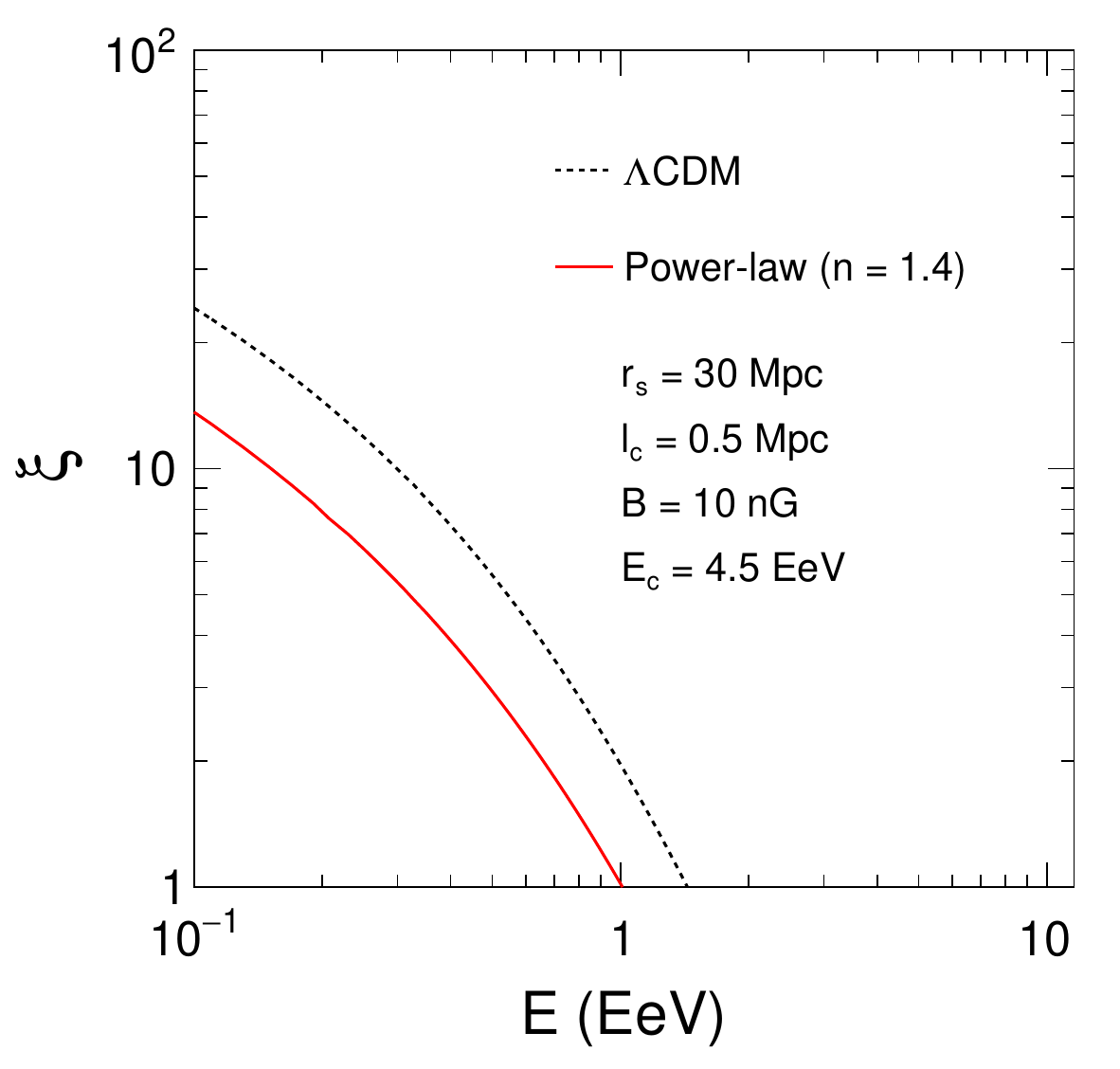}
\includegraphics[scale=0.25]{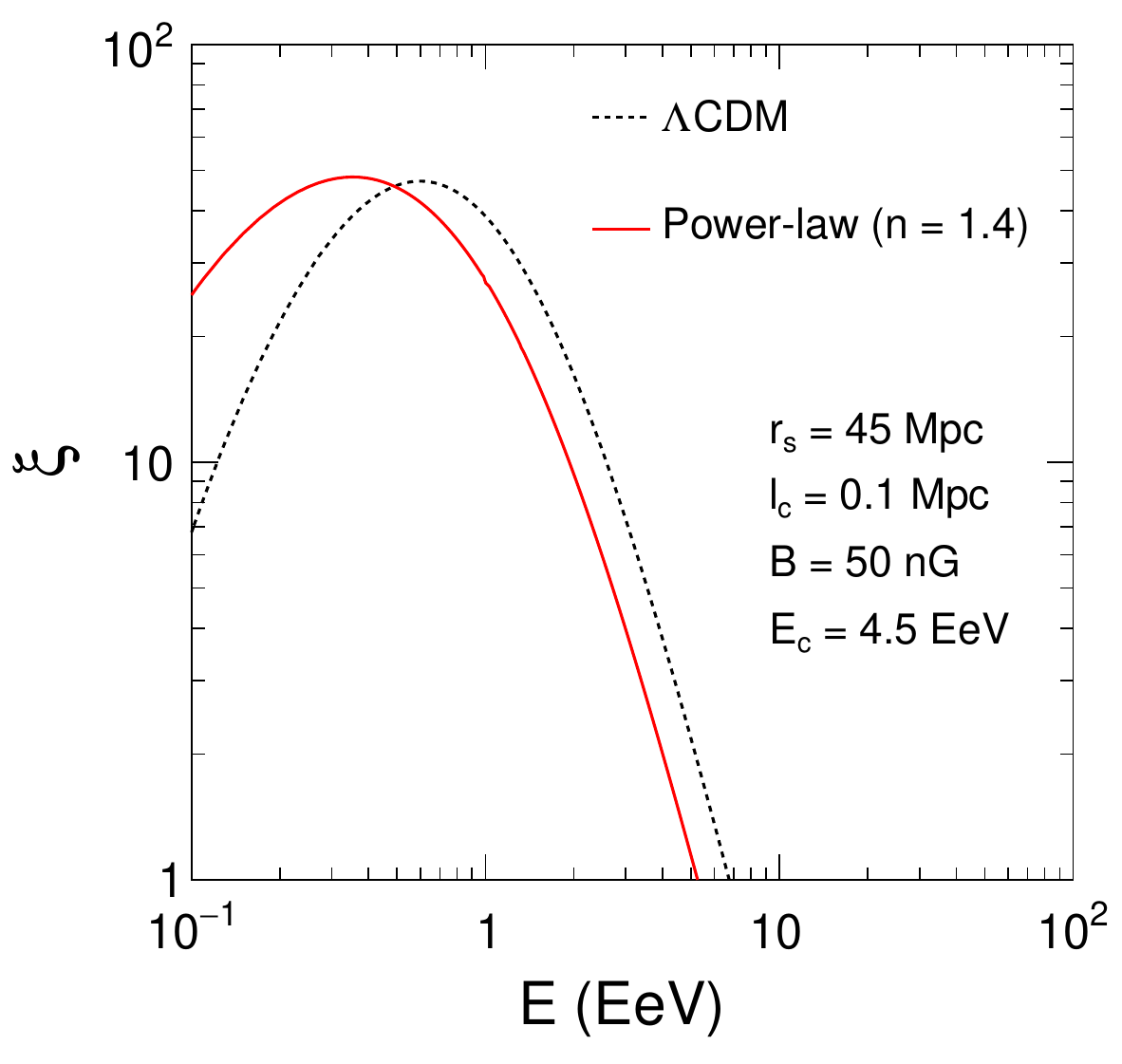}
\includegraphics[scale=0.25]{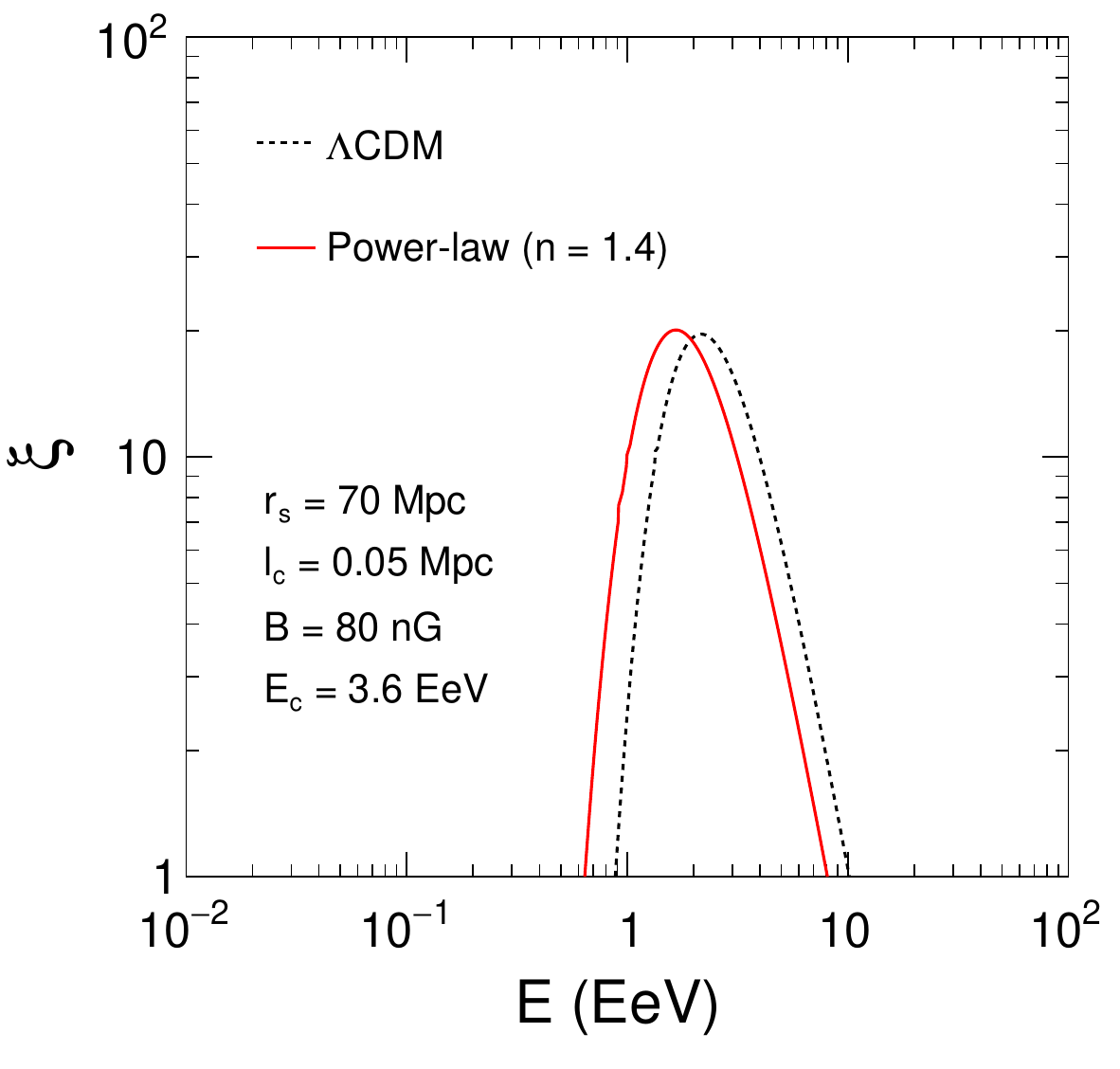}
}
\caption{Density enhancement $\xi$ as function of Energy $E$ for $\Lambda$CDM model (dotted line) and f(R) power law model (blue line). The corresponding Hubble parameters are 67.77 $\text{km}~ \text{s}^{-1}~ \text{Mpc}^{-1}$ and 68.4 $\text{km}~ \text{s}^{-1}~ \text{Mpc}^{-1}$ for the $\Lambda$CDM model and the power law model ($\beta = 1.4$).}
\label{fig1}
\end{figure}

The enhancement of density in Eq. \eqref{enhance} is plotted in
Fig. \ref{fig1}. Here, we vary the source distance $r_s$ from 30 Mpc to 70 Mpc, coherence length $l_c$ from 0.05 Mpc to 0.5 Mpc, and the strength of the magnetic field B is from 10 nG to 80 nG. In this figure, the black dotted line represents the standard $\Lambda$CDM model while the red line represents the f(R) gravity power law model.

\begin{figure}[h!]
\centerline{
\includegraphics[scale=0.25]{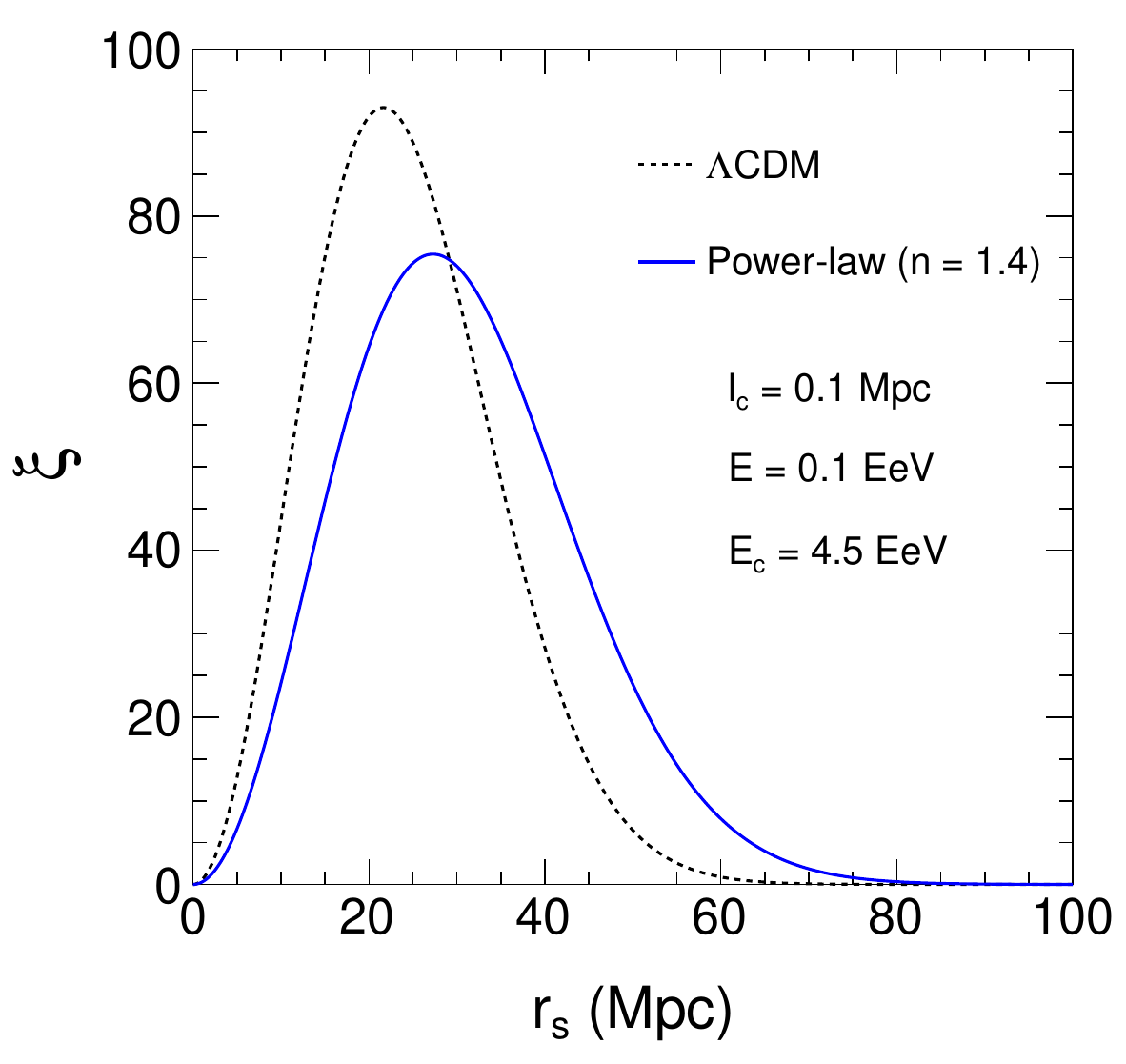}
\includegraphics[scale=0.25]{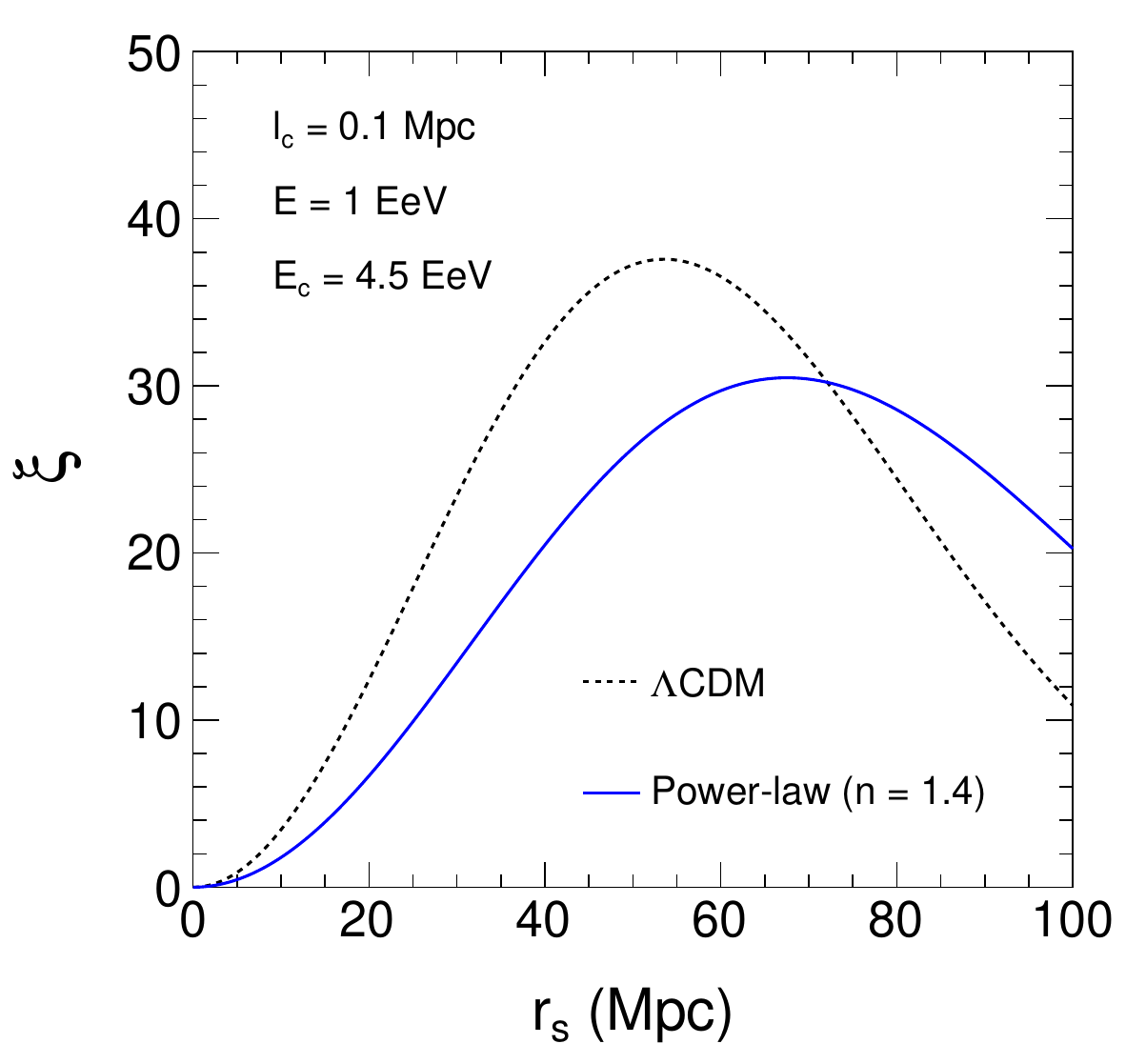}
\includegraphics[scale=0.25]{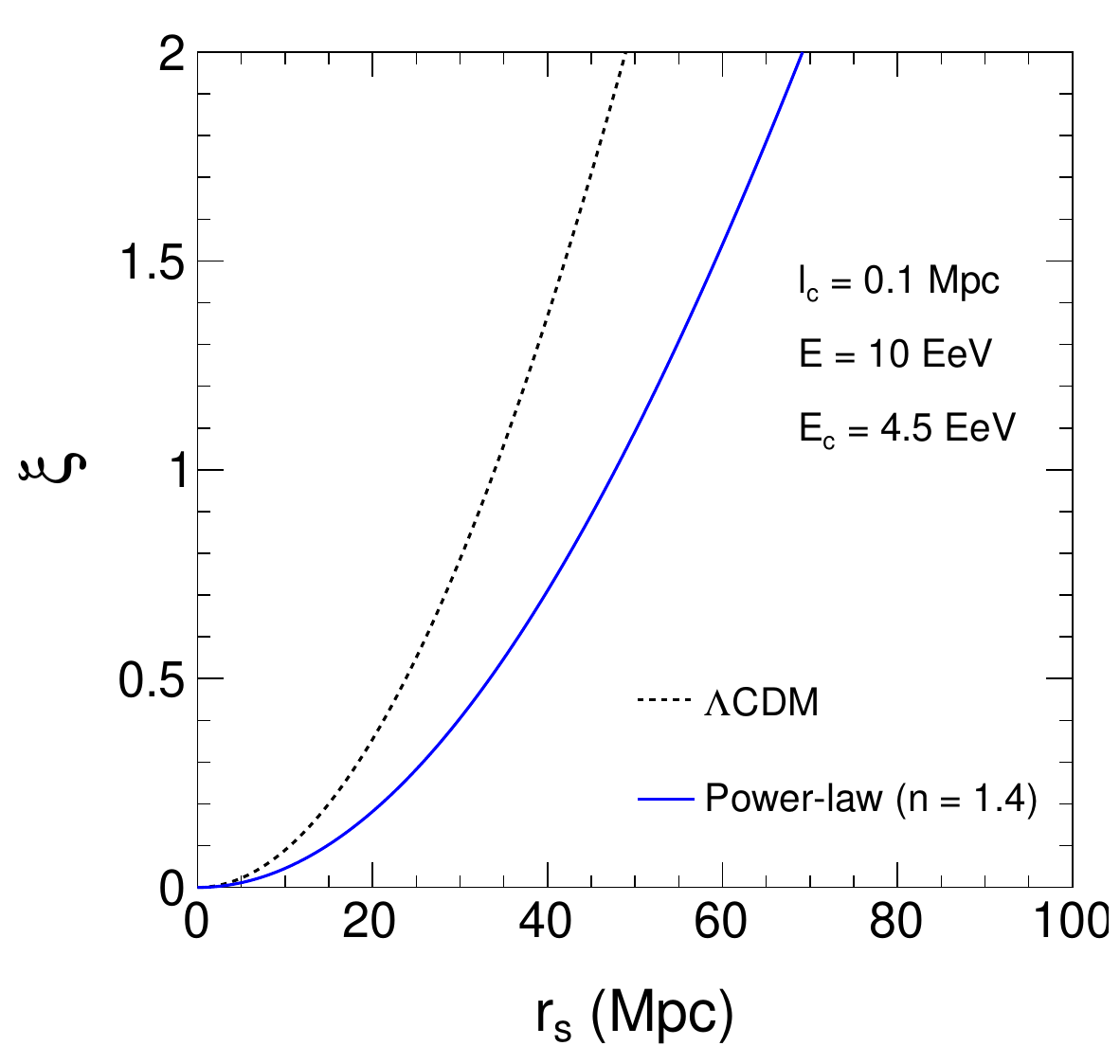}
}
\caption{Density enhancement $\xi$ as function of source distance $r_s$ for $\Lambda$CDM model (dotted line) and f(R) power law model (red line). The corresponding Hubble parameters are 67.77 $\text{km}~ \text{s}^{-1}~ \text{Mpc}^{-1}$ and 68.4 $\text{km}~ \text{s}^{-1}~ \text{Mpc}^{-1}$ for the $\Lambda$CDM model and the power law model ($\beta = 1.4$).}
\label{fig2}
\end{figure}

\begin{figure}[h!]
\centerline{
\includegraphics[scale=0.26]{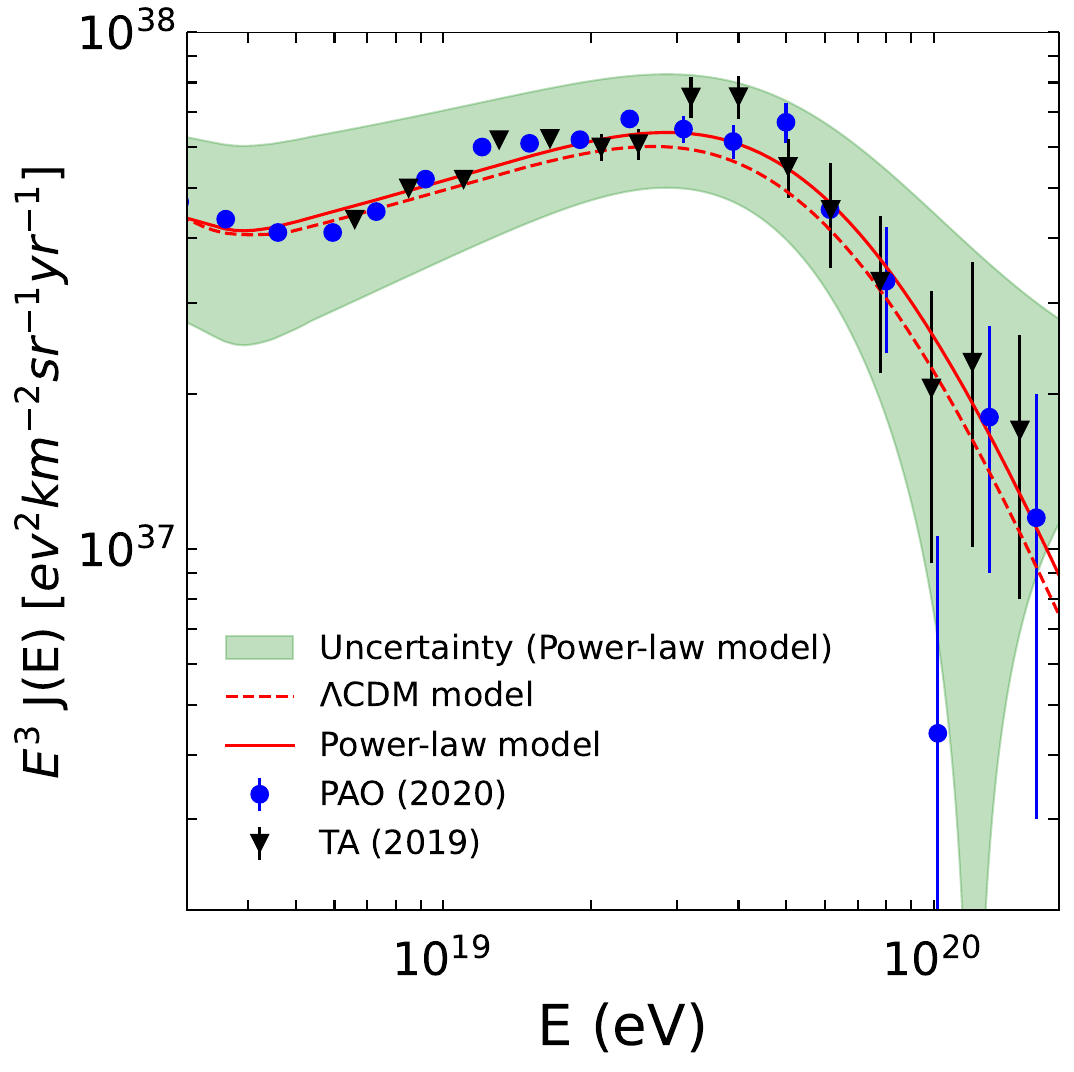}
}
\caption{Calculated flux for UHECRs for $\Lambda$CDM model (dotted line)  and f(R) power law model (solid line) and comparison with PAO and TA results with the uncertainties. The corresponding Hubble parameters are 67.77 $\text{km}~ \text{s}^{-1}~ \text{Mpc}^{-1}$ and 68.4 $\text{km}~ \text{s}^{-1}~ \text{Mpc}^{-1}$ for the $\Lambda$CDM model and the power law model ($\beta = 1.4$).}
\label{fig3}
\end{figure}
In the first panel of Fig. \ref{fig1}, the density enhancement of UHECR in the f(R) model is low in comparison to $\Lambda$CDM model whether in low energy or high energy region. In the next panel, the f(R) power-law model gives higher enhancement than $\Lambda$CDM model in the lower energy region ($\leq$ 1 EeV). In the third panel of Fig. \ref{fig1}, both cosmological models show similar behaviour. At about 1-2 EeV, the power law model dominates while after that $\Lambda$CDM model does the job. For more clarification, we plot 
 $xi$ with respect to $r_s$ in Fig. \ref{fig2}. Here, we can see that the peak of the enhancement is higher in the $\Lambda$CDM model while the power-law model presents a better distribution of enhancement. And we can conclude that in the lower energy region the shorter source distance dominates while in the higher energy region, enhancement takes place for the distance that is far away from the source.
In Fig. \ref{fig3}, We calculate the flux of UHECR for $\Lambda$CDM model and power law model. We compare our results with the combined observational results of Pierre Auger observatory \cite{augerprd2020} and Telescope array \cite{ta2019}. We can see that both models show a very good agreement with the observational results. At below 4 EeV, both models predict a similar kind of result. After that range, the power-law model predicts a slightly higher flux than the standard $\Lambda$CDM one. The green highlighted region denotes the uncertainties of our considered f(R) model and it is nicely confined within the observational data range with the error bars. For  the goodness of the fitting with
experimental data, we perform $\chi^2$ test that is given as
\begin{equation}
    \chi^2 = \sum_{i} \frac{(J^i_\text{th}-{J}^i_\text{obs})^2}{\sigma^2},
\end{equation}
where $J^i_\text{th}$ and $J^i_\text{obs}$ are the $i$th value of theoretical and observational value of flux respectively, and $\sigma$ denotes the standard deviation. With the PAO data, we get 2.79 and 2.52 for the $\Lambda$CDM and the power law model, while with the TA data, we get 2.84 and 2.74 respectively.

\section{Summary}  \label{IV}
We discuss the diffusive behaviour of UHECRs propagation in the standard $\Lambda$CDM model and f(R) power-law model. We calculate the density enhancement for both the models by varying the parameters such as source distance $r_s$, coherence length $l_c$, and strength of the magnetic field B. We see that our considered f(R) power law model shows a good agreement with the standard $\Lambda$CDM results. For prediction of the energy spectrum, we also calculate $\textrm{E}^3$ magnified flux for the f(R) model, and for comparison, we also implement the result of the $\Lambda$CDM model. For verification of our calculations, we compare our results for both models with the observational data of the Pierre Auger Observatory (PAO) and Telescope Array (TA). Both models show a good agreement with the observational data and it justifies the validity of our considered f(R) model. However, it should be noted that at this stage we assume our findings are for protons only, and more work needs to be done for other possible explanations. 

\section*{Acknowledgement}
SPS acknowledges Prof. U. D. Goswami, Department of Physics, Dibrugarh University for some useful discussions.


\section*{References}
\begin{thebibliography}{99}
\bibitem{Hess} V. F. Hess, Phys. Z. \textbf{13}, 1084 (1912).

\bibitem{harari} D. Harari, S. Mollerach, E. Roulet, Phys. Rev. D \textbf{89}, 123001 (2014).
\bibitem{molerach} S. Mollerach, E. Roulet, Phys. Rev. D \textbf{99}, 103010 (2019).
\bibitem{berezinskyGK} V. Berezinsky, A. Z. Gazizov, O. Kalashev, Astropart. Phys., \textbf{84}, 52 (2016).
\bibitem{berezinskyGG} V. Berezinsky, A. Z. Gazizov, S. I. Grigorieva, arXiv:astro-ph/0210095.
\bibitem{Bhattacharjee}P. Bhattacharjee, G. Sigl, Phys. Rept. \textbf{327} (2000).
\bibitem{Olinto} A. V. Olinto, Phys. Rept. \textbf{333} (2000).
\bibitem{s.mollerach} S. Mollerach, E. Roulet, Phys. Rev. D \textbf{105} 063001 (2022).
\bibitem{Supanitsky} A. D. Supanitsky, JCAP \textbf{04}, 046 (2021).
\bibitem{mollerach2013}S. Mollerach, E. Roulet, JCAP \textbf{10}, 013 (2013). 
\bibitem{berezinkyGre} V. Berezinsky, A. Z. Gazizov, Astrophys. J. \textbf{643}, 8 (2006).
\bibitem{berezinski_four_feat} V. Berezinsky, A. Z. Gazizov, S. I. Grigorieva, Phys. Rev. D \textbf{74}, 043005 (2006).
\bibitem{gogoi} D. Gogoi and U. D. Goswami ,IJMP D \textbf{31}, 225048 (2022).
\bibitem{aagnamin} N. Aghanim et al., Planck 2018 results (Planck Collaboration), A \& A 641, A6 (2020).
\bibitem{nakamura} K. Nakamura and Particle Data Group, Review of Particle Physics, J. Phys. G: Nucl. Part. Phys. 37, 075021(2010).
\bibitem{augerprd2020}A. Aab et al. (Pierre Auger
Collaboration) Phys. Rev. D \textbf{102}, 062005 (2020).
\bibitem{ta2019} D. Ivanov (Telescope Array Collaboration),
Proc. Sci. ICRC 2019, \textbf{298} (2019).


\end{thebibliography}
\end{document}